\documentclass[preprint,preprintnumbers,amsmath,amssymb]{revtex4}

\usepackage{graphicx}
\usepackage{dcolumn}
\usepackage{bm}

\begin{document}

\title{Long-Time Behavior of Velocity Autocorrelation Function 
for Interacting Particles
in a Two-Dimensional Disordered System}

\author{Tatsuro Yuge}
\email{yuge@ASone.c.u-tokyo.ac.jp}
\author{Akira Shimizu}
\email{shmz@ASone.c.u-tokyo.ac.jp}
\affiliation{Department of Basic Science, 
University of Tokyo, 3-8-1 Komaba, Meguro-ku, Tokyo 153-8902, Japan}

\date{\today}

\begin{abstract}
The long-time behavior of the velocity autocorrelation function (VACF) 
is investigated by the molecular dynamics simulation 
of a two-dimensional system 
which has both a many-body interaction and a random potential.
With strengthening the random potential 
by increasing the density of impurities,
a crossover behavior of the VACF 
is observed from a positive tail, which is proportional to $t^{-1}$,
to a negative tail, proportional to $-t^{-2}$.
The latter tail exists even when the density of particles
is the same order as the density of impurities.
The behavior of the VACF in a nonequilibrium steady state is also studied.
In the linear response regime the behavior is similar to that in the equilibrium state,
whereas it changes drastically in the nonlinear response regime.
\end{abstract}

\maketitle

Transport coefficients are related to time integrals 
of the time correlation functions of appropriate quantities.
Sufficiently fast decay of the correlation functions is therefore necessary 
for convergence of the transport coefficients,
and it had been expected that the correlation functions 
are exponentially decaying functions in time.
There are, however, certain systems 
in which correlation functions exhibit power-law decay at long times,
which is called the long-time tail
\cite{LLP2,Murakami,OYI,SYI,AW1,EHL1,DC1,ZB,
Kawasaki1,PomeauResibois,KBS,EW,Bruin1,AA1,AA2,GLY1,EMDB1,WBSM,HF}.
In heat conduction systems such as 
nonlinear lattices, hard-core particles, and Lennard-Jones systems,
the long-time tails of the autocorrelation functions of the heat flux
and the resultant system-size dependence 
of the thermal conductivity have been observed numerically 
\cite{LLP2,Murakami,OYI,SYI}.

The long-time tails have also been observed 
for the velocity autocorrelation function (VACF) of a moving particle,
which determines the self-diffusion coefficient $D$,
in particle transport systems
\cite{AW1,EHL1,DC1,ZB,Kawasaki1,
PomeauResibois,KBS,EW,Bruin1,AA1,AA2,GLY1,EMDB1,WBSM,HF}.
One of such systems is the hard-core fluid 
where the long-time tail of the VACF is positive and proportional 
to $t^{-d/2}$ 
(which we call the fluid-type tail) 
\cite{AW1,EHL1,DC1,ZB,Kawasaki1,PomeauResibois,KBS}.
Here $d$ is the dimension of the system.
Hence $D$ is divergent in the two-dimensional fluid,
which may be observed as 
logarithmic dependence of $D$ 
on the system size.
Another system is the Lorentz model,
which describes the motion of a single point particle 
in a disordered system where immobile scatterers are randomly arrayed.
The long-time tail of the VACF in the Lorentz model 
is known to be negative and proportional to $-t^{-(d/2+1)}$ 
(which we call the Lorentz-type tail) 
\cite{EW,Bruin1,AA1,AA2,GLY1,EMDB1}.
Although this tail does not lead to divergence of $D$ 
even for $d=2$,
it might induce a system-size dependent term in $D$
which decays only algebraically with increasing the system size.

Although in the fluid system only a many-body interaction is present 
whereas in the Lorentz model only a random potential 
(which immobile scatterers generate) is present,
there are many physical systems which have both of them. 
A typical one of such systems is the electric conduction system,
where interacting particles correspond to electrons and 
a random potential is generated by impurities and defects.
At room temperature,
two-dimensional electron systems in semiconductors
are well described as two-dimensional {\em classical} systems
because the room temperature is smaller than 
the quantum level spacing in the confining ($z$) direction 
and larger than the Fermi energy in the $x$-$y$ plane \cite{2DEG}.
It is widely known from experimental results that $D$ 
of such systems
is well-defined, independent of the sample size,
to a good approximation.
This fact suggests that 
the {\em coexistence} of a many-body interaction (among electrons)
and a random potential 
would change the long-time behavior of the VACF 
from the above-mentioned theoretical results 
which assume existence of only {\em either one} of them.

In the present paper,
using the model of a classical electric conduction system proposed by us 
\cite{YIS},
we compute the VACF in a two-dimensional system
which has both a random potential (impurity scatterings)
and a many-body [electron-electron ({\it e-e})] interaction,
and study how the long-time behavior of the VACF changes
with varying the densities of impurities and electrons
\cite{note1}.
We also investigate the VACF in nonequilibrium steady states,
and observe considerable changes in the nonlinear response regime.

First, we analyze the equilibrium state.
For this purpose we use a simplified version of our model \cite{YIS}.
It consists of two types of classical hard disks
in a two-dimensional square box, the linear size of which is $L$.
One type, which we call impurity, is immobile.
Its radius and number density are denoted 
by $R_i$ and $n_i$, respectively.
The other type, which we call electron, moves freely 
till it collides with another electron or with an impurity.
The mass, radius, and number density of the electrons 
are denoted by $m_e$, $R_e$, and $n_e$, respectively.
We can change the strength of the random potential 
and the frequency of {\it e-e} collisions 
by varying the values of $n_i$ and $n_e$, respectively.
The model without the impurities corresponds to the hard-disk fluid 
whereas the model without the {\it e-e} interaction
corresponds to the (non-overlapping) Lorentz model.

Using the event-driven molecular dynamics (MD) simulation, 
we study the long-time behavior of the VACF,
$C(t)\equiv \langle \mbox{\boldmath$v$}(t)\cdot\mbox{\boldmath$v$}(0)\rangle / d$,
of an electron in an equilibrium state $(d=2)$.
The data of the VACF shown below are computed as follows:
First we calculate the VACF for each electrons 
and take the average over the electrons in the system 
for each configuration of the impurities.
Then we average the results over five configurations of the impurities.
The errorbars are the standard deviations 
among the results of the five configurations.

We set $m_e$, $R_e$, and the temperature $k_{\mathrm{B}} T$ to be unity,
and take $L$ and $R_i$ to be $L=480$ and $R_i=0.4$ in these units.
The boundary conditions are set to be periodic 
in both directions.
The impurities are distributed randomly under the restriction 
that the distances among them are larger than $2(R_e + R_i)$.
The initial positions of the electrons are so randomly arranged 
as not to overlap with the other disks,
and their initial velocities are given 
by the Maxwell distribution of unit temperature.
We calculate the VACF in the following two cases:
(i) when we fix $n_e$ and change $n_i$, 
i.e.,
we introduce the random potential into the system 
where only the {\it e-e} interaction is present, 
(ii) when we fix  $n_i$ and change $n_e$, 
i.e.,
we introduce the {\it e-e} interaction into the system 
where only the random potential is present.

\begin{figure}
\begin{center}
\includegraphics[width=0.8\linewidth]{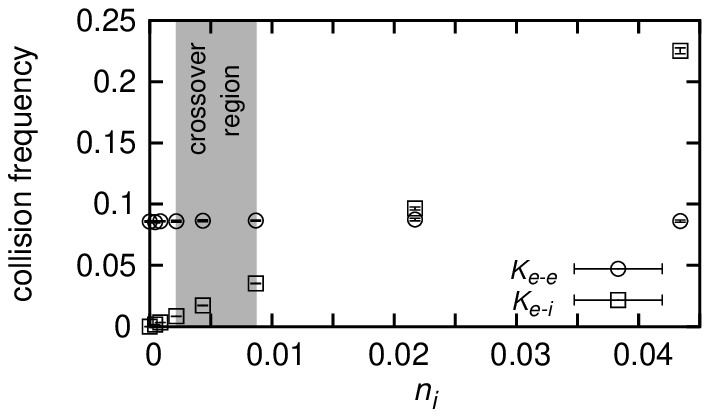}
\end{center}
\caption{\label{fig:kaisu_e5000}
Collision frequencies per unit time,
$K_{e\mbox{-}e}$ (circles) and $K_{e\mbox{-}i}$ (squares), plotted against $n_i$.
The value of $n_e$ is fixed to 0.0217.
The sizes of most of the errorbars are smaller than the sizes of the symbols.
The region in gray shows the crossover region 
between the fluid-type and Lorentz-type tails.
}
\end{figure}

\begin{figure}
\begin{center}
\includegraphics[width=0.8\linewidth]{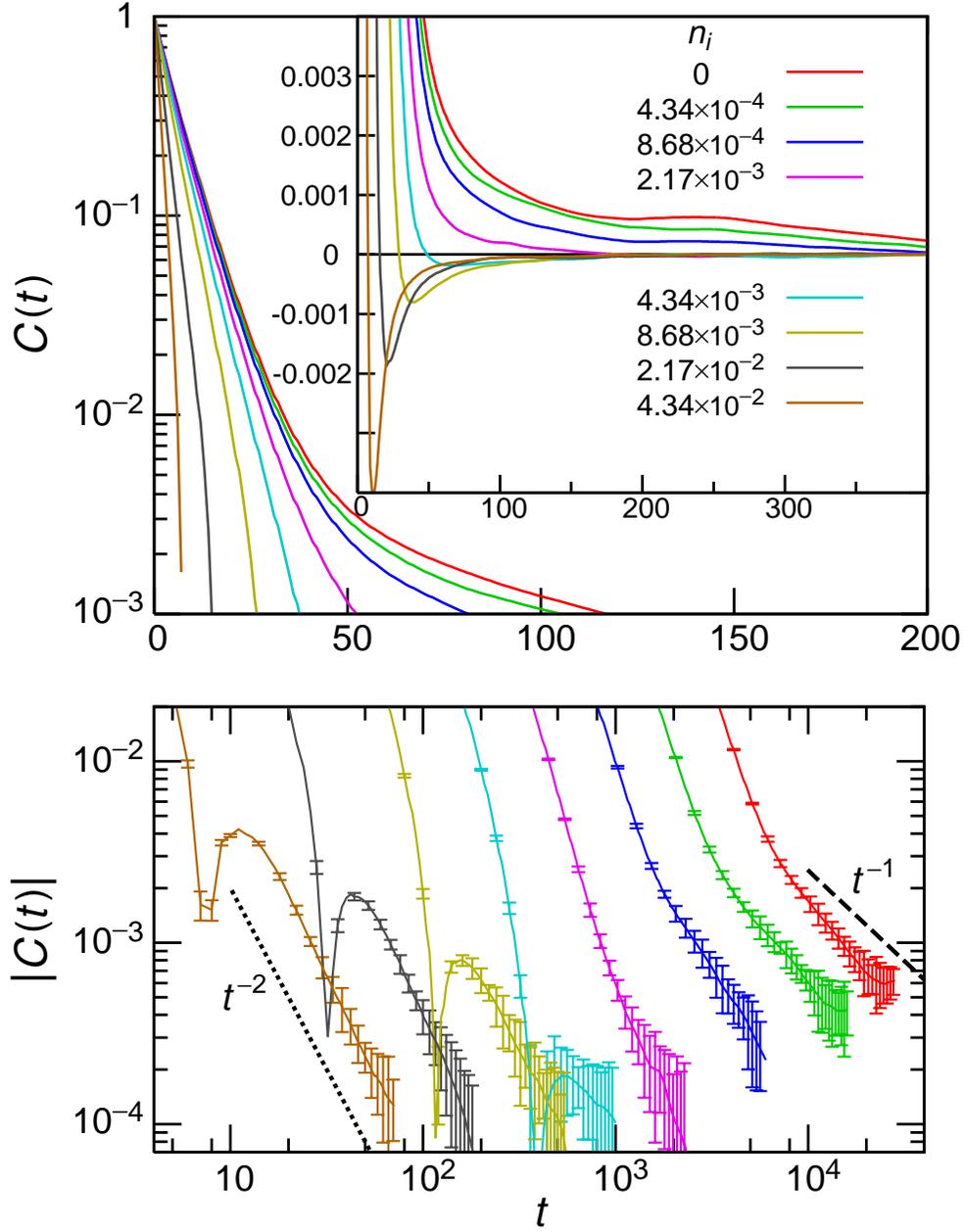}
\end{center}
\caption{\label{fig:e5000}
The VACF plotted against $t$, for various values of $n_i$.
The parameters are the same as those in Fig.~\ref{fig:kaisu_e5000}.
Top: A semilog plot around initial times.
Inset of the top: A magnified plot on linear scales.
Bottom: A double logarithmic plot of the absolute value of the VACF. 
In the bottom, for clarity, the data of neighboring parameters 
have been shifted along the time axis by a factor of two.
The dashed and dotted lines are guides to the eye,
which are proportional to $t^{-1}$ and $t^{-2}$, respectively.}
\end{figure}
In case (i), we calculate the VACF of the system
with $n_e$ fixed to 0.0217 and $n_i$ varied from 0 to 0.0434.
In this case, the reduced electron density is $n_e (R_e+R_e)^2 = 0.0868$.
In Fig.~\ref{fig:kaisu_e5000} we show the {\it e-e} and {\it e-i} collision frequencies 
(denoted by $K_{e\mbox{-}e}$ and $K_{e\mbox{-}i}$, respectively).
Here, $K_{e\mbox{-}e}=$ (number of the {\it e-e} collisions per unit time)$/n_e L^2$
and $K_{e\mbox{-}i}=$ (number of the {\it e-i} collisions per unit time)$/n_e L^2$.
From this result the duration time $\tau_{e\mbox{-}e}$ $(\equiv 1/K_{e\mbox{-}e})$ 
of the {\it e-e} collisions
is evaluated to be about 12 when $n_i =0$.
The numerical results of the VACF are shown in Fig.~\ref{fig:e5000}.
We observe that $C(0)\simeq 1= k_{\mathrm{B}} T/m_e$,
implying the law of equipartition of energy,
and that the VACF decays exponentially at initial times 
and then decays algebraically at longer times.
There exists the positive long-time tail 
for $n_i \lesssim 2.17\times 10^{-3}$.
The exponent is about $-1$ at $n_i =0$,
which reproduces the result for the hard-disk fluid
\cite{AW1,EHL1,DC1,ZB,Kawasaki1,PomeauResibois,KBS}.
With increasing $n_i$ this fluid-type tail becomes weaker 
(i.e., the decay becomes faster).
For $n_i \gtrsim 4.34\times 10^{-3}$ it disappears,
and instead, the negative long-time tail appears.
The VACF decays as $-t^{-2}$ for $n_i \gtrsim 10^{-2}$,
which is the same result as the Lorentz model
\cite{EW,Bruin1,AA1,AA2,GLY1,EMDB1}.
This observation suggests that the Lorentz-type tail 
exists even when $\tau_{e\mbox{-}e}$ is the same order as 
$\tau_{e\mbox{-}i}=1/K_{e\mbox{-}i}$, the duration time of the {\it e-i} collisions.

\begin{figure}
\begin{center}
\includegraphics[width=0.8\linewidth]{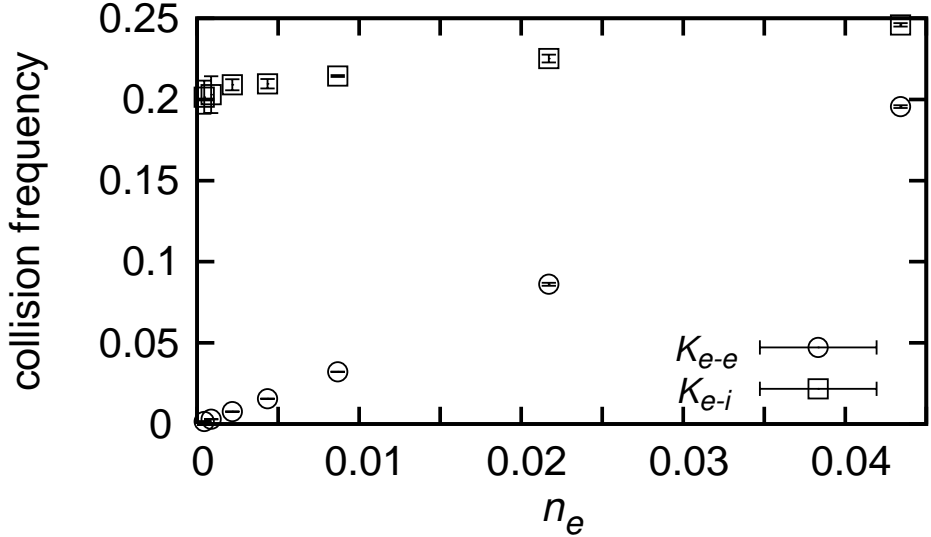}
\end{center}
\caption{\label{fig:kaisu_i10000}
Collision frequencies per unit time,
$K_{e\mbox{-}e}$ (circles) and $K_{e\mbox{-}i}$ (squares), plotted against $n_e$.
The value of $n_i$ is fixed to 0.0434.
The sizes of most of the errorbars are smaller than the sizes of the symbols.
The crossover region would exist at larger values of $n_e$ 
than the range shown here.
}
\end{figure}

\begin{figure}
\begin{center}
\includegraphics[width=0.8\linewidth]{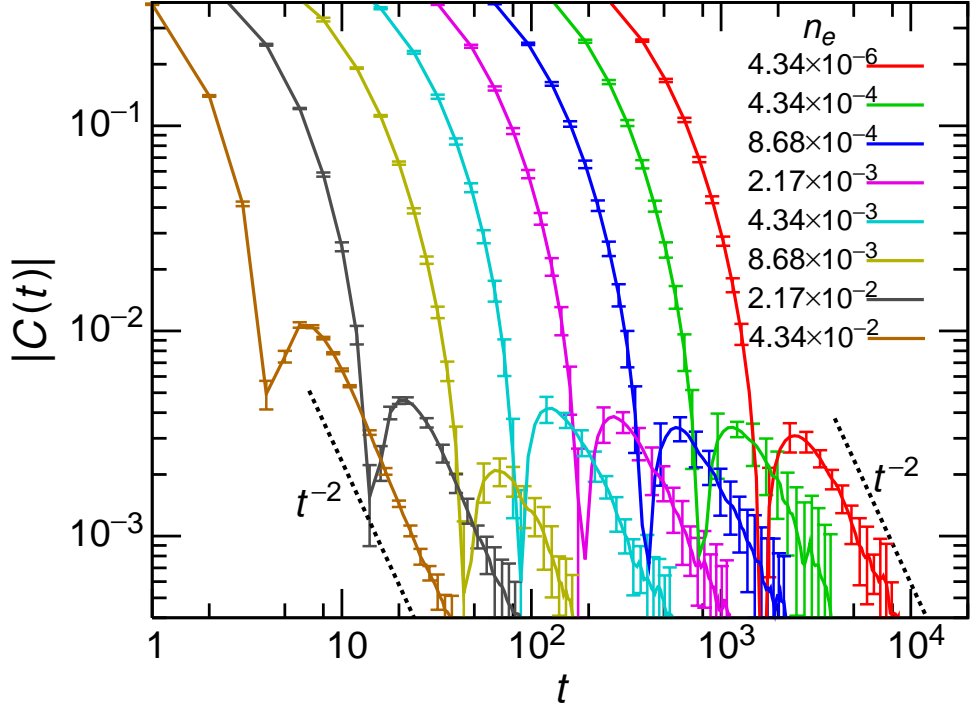}
\end{center}
\caption{\label{fig:i10000}
A double logarithmic plot of the absolute value of the VACF,
plotted against $t$, for various values of $n_e$.
The parameters are the same as those in Fig.~\ref{fig:kaisu_i10000}.
The red line ($n_e = 4.34\times 10^{-6}$) 
represents the case of no {\it e-e} interaction,
which corresponds to the Lorentz model.
For clarity, the data of neighboring parameters 
have been shifted along the time axis by a factor of two.
The dotted lines are guides to the eye,
which are proportional to $t^{-2}$.
}
\end{figure}
To see this fact more clearly,
we compute the VACF for case (ii),
with $n_i$ fixed to 0.0434 and $n_e$ varied from $4.34\times 10^{-6}$ 
(i.e., a single electron in the system,
corresponding to the system without the {\it e-e} interaction) to 0.0434.
In this case, the reduced impurity density is $n_i (R_e+R_i)^2 = 0.0851$,
and $K_{e\mbox{-}e}$ and $K_{e\mbox{-}i}$ are shown 
in Fig.~\ref{fig:kaisu_i10000}.
From this result $\tau_{e\mbox{-}i}\simeq 5$ 
when the {\it e-e} interaction is absent.
We show the numerical results of the VACF in Fig.~\ref{fig:i10000}.
It is seen that at long times the VACF decays as $-t^{-2}$ at $n_e=4.34\times 10^{-6}$,
which corresponds to the Lorentz model,
and that this Lorentz-type tail also exists for larger values of $n_e$.
Thus the Lorentz-type tail appears 
even when $\tau_{e\mbox{-}e} \sim \tau_{e\mbox{-}i}$.
(see Fig.~\ref{fig:kaisu_i10000}).

According to the phenomenological approaches
\cite
{EHL1,Kawasaki1,PomeauResibois,KBS,GLY1,EMDB1},
the origin of the long-time tails is the coupling of hydrodynamic modes in the systems.
In the fluid the conservation of both the momentum and particle number 
contributes to the hydrodynamic modes 
\cite{EHL1,PomeauResibois,KBS}.
In the Lorentz model, by contrast,
only the particle-number conservation contributes to the mode,
which couples with the static mode generated 
by the configuration of the impurities \cite{EMDB1}.
Although the absence of the interaction between particles 
(the {\it e-e} interaction) was assumed in ref.~17, 
our results suggest that a similar mode-coupling theory would be valid 
whether or not the interaction between particles is present
if a system has the dynamical mode from particle-number conservation 
and some static modes couple with it.

According to this interpretation,
the crossover of the long-time behavior 
from the fluid-type tail to the Lorentz-type one 
with increasing $n_i$ (Fig.~\ref{fig:e5000})
would be explained as follows.
To treat the electron system as a fluid,
it is necessary that the {\it e-e} collisions occur more frequently
than the {\it e-i} collisions.
That is, $\tau_{e\mbox{-}i} \gtrsim M_c \tau _{e\mbox{-}e}$ 
is a necessary condition for regarding the velocity field to be hydrodynamic,
where $M_c~ (\gtrsim 1)$ is a crossover ratio of the collision frequencies,
which would be almost independent of the densities and the system size.
Therefore, when $\tau_{e\mbox{-}i}/\tau_{e\mbox{-}e} \gg M_c$
both types of the long-time tails are present,
and we observe the fluid-type one manifestly 
because it is stronger than the Lorentz-type one.
With increasing $n_i$, $\tau_{e\mbox{-}i}$ becomes smaller, 
and when $\tau_{e\mbox{-}i}/\tau_{e\mbox{-}e} \ll M_c$
the fluid-type tail disappears 
because the hydrodynamic description of the velocity field is not valid.
The Lorentz-type tail, by contrast, survives even in this regime 
due to the particle-number conservation.

In Fig.~\ref{fig:e5000} 
the crossover takes place around
$2.17 \times 10^{-3} \lesssim n_i \lesssim 8.68 \times 10^{-3}$.
Thus from the corresponding values of $\tau_{e\mbox{-}i}$ and $\tau_{e\mbox{-}e}$
in Fig.~\ref{fig:kaisu_e5000},
we evaluate $M_c$ to satisfy 
\begin{equation}
2\lesssim M_c \lesssim 10.
\label{M}
\end{equation}
We can translate $M_c$ into the crossover value of the 
density ratio as
\begin{equation}
(n_e/n_i)_c=M_c\sigma_{e\mbox{-}i}/\sigma_{e\mbox{-}e}
\label{tau}\end{equation}
if we assume the approximate relations
$\tau_{e\mbox{-}i} \sim 1/v_e n_i \sigma_{e\mbox{-}i}$ 
and $\tau_{e\mbox{-}e} \sim 1/v_e n_e \sigma_{e\mbox{-}e}$.
Here, $v_e$ is the thermal velocity of an electron,
and $\sigma_{e\mbox{-}i}$ and $\sigma_{e\mbox{-}e}$ 
are the cross sections of the {\it e-i} and {\it e-e} scatterings, respectively.
Since $\sigma_{e\mbox{-}i}/\sigma_{e\mbox{-}e}=(R_e+R_i)/2R_e$ 
for hard-disk interactions, 
eq.~(\ref{M}) is translated as
$2 \lesssim (n_e/n_i)_c \lesssim 7$.
These values are consistent with Fig.~\ref{fig:e5000}
(where $n_e$ is fixed to $2.17 \times 10^{-2}$),
which shows the validity of eq.~(\ref{tau}).
In Fig.~\ref{fig:i10000} ($n_i=4.34 \times 10^{-2}$),
$n_e/n_i$ ranges from $10^{-3}$ to $1$ 
in our calculation,
which is smaller than $(n_e/n_i)_c$.
Therefore, the fact that 
we have observed only the Lorentz-type tail in Fig.~\ref{fig:i10000}
is also consistent with the above interpretation.

\begin{figure}
\begin{center}
\includegraphics[width=0.8\linewidth]{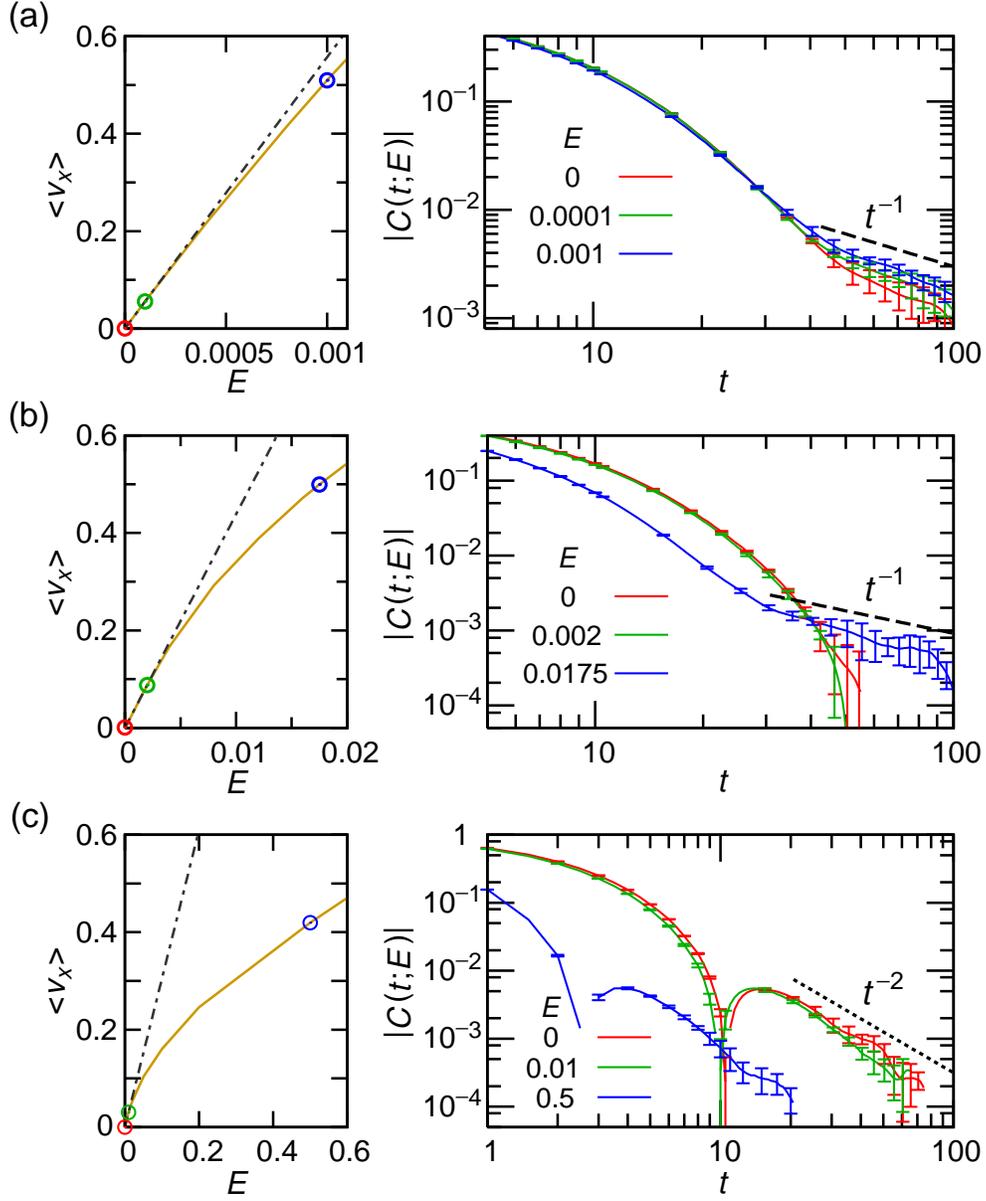}
\end{center}
\caption{\label{fig:noneq}
Left graphs: The mean velocities of an electron versus electric fields.
The broken lines represent the slopes of the linear response coefficients.
Right graphs: The absolute values of the VACF
for steady states under various values of the electric field,
plotted against $t$.
The data in red, green and blue are the results 
for the equilibrium state,
the linear response regime 
and the nonlinear response regime, respectively.
The dashed and dotted lines are guides to the eye,
which are proportional to $t^{-1}$ and $t^{-2}$, respectively.
The number densities of the impurities are 
(a) $n_i = 0 ~[n_e/n_i > (n_e/n_i)_c]$,
(b) $n_i = 0.00213 ~[n_e/n_i \sim (n_e/n_i)_c]$, and
(c) $n_i = 0.0267 ~[n_e/n_i < (n_e/n_i)_c]$.
Unlike Figs.~\ref{fig:e5000} and \ref{fig:i10000},
we have not shifted any of the data along the horizontal axes.}
\end{figure}
Finally, let us examine the long-time tail of the VACF in nonequilibrium steady states.
In order to realize a nonequilibrium steady state,
we use our original model \cite{YIS}.
That is, in addition to the electrons and impurities 
there exist another type of moving disks
(which we call phonons) in a rectangular box of
size $L_x\times L_y$.
The mass and radius of the phonons are $m_p=1.0$ and $R_p=1.0$, respectively.
In the $x$-direction we apply an electric field $E$ to the electrons 
to induce the electric current,
and the boundary condition in this direction is periodic.
The boundaries in the $y$-direction are potential walls for the electrons,
and thermal walls for the phonons,
which simulate the heat transfer outside the system (conductor) 
to keep it in a steady state.
The interaction between any pair of the disks is taken to be Hertzian.
The equations of motion are solved by the Gear's predictor-corrector method.
As shown in ref.~21 and in the left graphs in Fig.~\ref{fig:noneq}
(in which the charge of an electron is set to be unity),
this model has well-defined nonequilibrium steady states
even in the nonlinear response regime.
We calculate the VACF defined by
\begin{eqnarray}
C(t;E)\equiv 
\frac{\bigl\langle (v_x(t)-\langle v_x \rangle) (v_x(0)-\langle v_x \rangle) \bigr\rangle}
{\bigl \langle (v_x -\langle v_x \rangle)^2 \bigr\rangle}
\end{eqnarray}
in a nonequilibrium steady state,
where $v_x$ is the component of electron's velocity parallel to $E$.
We set 
$L_x=750$ and $L_y=125$,
and fix the number densities of the electrons and phonons to be 
$n_e = 0.0160$ and $n_p = 0.00533$, respectively.
From the results for the equilibrium state ($E=0$),
we find that the crossover of the VACF with increasing $n_i$ is observed 
even when the phonons are present.
Then we compute and compare the VACFs 
in the equilibrium state, the linear response regime and the nonlinear response regime,
for various values of $n_e/n_i$.
The right graphs in Fig.~\ref{fig:noneq} depict the results.
It is seen that 
in the linear response regime the behavior of the VACF is almost the same 
as in the equilibrium state.
In the nonlinear response regime, by contrast,
we observe drastic changes in the behavior.
For $n_e/n_i \gtrsim (n_e/n_i)_c$ [Figs.~\ref{fig:noneq}(a) and \ref{fig:noneq}(b)] 
the fluid-type tail is enhanced 
as the electric field becomes larger.
For $n_e/n_i \sim (n_e/n_i)_c$ [Fig.~\ref{fig:noneq}(b)]
in particular,
although we cannot observe the fluid-type tail clearly either
in the equilibrium state or in the linear response regime,
it appears manifestly in the nonlinear response regime.
For $n_e/n_i < (n_e/n_i)_c$ [Fig.~\ref{fig:noneq}(c)] 
the Lorentz-type tail appears from shorter times than in the equilibrium state.
We think that this is partly because $\tau_{e\mbox{-}i}$ becomes shorter 
with increasing $E$ \cite{YIS}.

In summary,
using the MD simulation
we have studied the long-time behavior of the VACF 
for interacting electrons in a two-dimensional disordered system.
For the equilibrium state we have found that 
with increasing the density of the impurities 
the long-time tail crosses over 
from the positive tail proportional to $t^{-1}$ (the fluid-type tail)
to the negative tail proportional to $-t^{-2}$ (the Lorentz-type tail).
The Lorentz-type tail survives 
even when we increase the density of the electrons 
up to the same order as that of the impurities.
These results imply that in our two-dimensional system 
the diffusion constant does not diverge in the limit of infinite system size 
\cite{note2}.
There may remain, however, a small term, which depends weakly (algebraically)
on the system size.
We have also found that the crossover occurs 
when $\tau_{e\mbox{-}i}/\tau_{e\mbox{-}e} \sim M_c$ or 
$n_e/n_i \sim M_c\sigma_{e\mbox{-}i}/\sigma_{e\mbox{-}e}$,
where $2\lesssim M_c \lesssim 10$.

For nonequilibrium steady states,
although in the linear response regime 
the VACF behaves similarly to that in the equilibrium state,
its behavior changes considerably in the nonlinear response regime.
There the fluid-type tail is enhanced for smaller $n_i$
whereas the Lorentz-type tail appears from earlier times for larger $n_i$.

We expect that 
these summarized results would be insensitive to details of the systems,
and that elaborate experiments and
a kinetic theory with two-parameter ($n_e$ and $n_i$) density expansion
would support our results.
Furthermore, 
our estimate of $M_c$ would be useful 
for analyzing not only the long tails but also other physical phenomena
of many-particle systems
because one can 
judge whether the system 
can be described as a fluid.

\section*{Acknowledgment}
The authors acknowledge N. Ito for helpful discussions.
This work was supported by a grant from the Research Fellowships 
of the Japan Society for the Promotion of Science for Young Scientists (No. 1811579).


\begin{thebibliography}{50}
\bibitem{note1}
While our calculation is performed for rather low densities,
recent works indicate that the exponents of the long-time tails change 
for much higher densities of particles for the hard-sphere fluid model \cite{WBSM}
and the Lorentz model \cite{HF}.
\bibitem{note2}
Some Burnett coefficients may diverge as in the Lorentz model \cite{EMDB1}.
Its significance for trasport phenomena in real physical systems remains an open question.
\end{thebibliography}
\end{document}